%% file: main.tex
\documentclass[conference]{IEEEtran}
\IEEEoverridecommandlockouts
\usepackage{cite}
\usepackage{amsmath,amssymb,amsfonts}
\usepackage{algorithmic}
\usepackage{graphicx}
\usepackage{textcomp}
\usepackage{xcolor}
\def\BibTeX{{\rm B\kern-.05em{\sc i\kern-.025em b}\kern-.08em
    T\kern-.1667em\lower.7ex\hbox{E}\kern-.125emX}}

\usepackage{booktabs,array,colortbl,tabularx}
\usepackage{xcolor}
\usepackage{pifont}
\usepackage[bookmarks=false]{hyperref}
\usepackage{booktabs}
\usepackage{microtype}
\usepackage{xspace}
\usepackage{enumitem}
\usepackage{balance}
\usepackage{stfloats}
\usepackage{colortbl}
\usepackage{multicol}
\usepackage{multirow}
\usepackage{pifont}

\definecolor{PendingRed}{HTML}{BF2535}
\definecolor{ReferencePurple}{HTML}{7541A8}
\definecolor{CitationPink}{HTML}{AD3670}

\begin{document}
\title{\textit{ActECC}: \underline{A}ction \underline{C}hunk \underline{T}emporal Harnessing for \underline{E}dge-\underline{C}loud \underline{C}ollaborative VLA Inference System}
\author{\IEEEauthorblockN{Zihao Zheng$^{1, *}$, Zhihao Mao$^{1, *}$, Hui Huang$^{2}$, Sicheng Tian$^{3}$, Hangyu Cao$^{1}$, Ziyun Zhang$^1$, Chenyue Li$^1$, \\ Jiayu Chen$^1$, Maoliang Li$^{1}$, Hailong Zou$^{1}$, Guojie Luo$^{1}$, Xiang Chen$^{1, \dagger}$}
\IEEEauthorblockA{
$^{1}\ $Peking University \ $^{2} \ $Beijing Jiaotong University \
$^{3} \ $Beijing Normal University \\
$^{*} \ $Equal Contribution \ $^\dagger \ $Corresponding author: xiang.chen@pku.edu.cn}}

\maketitle

% ================================================================
% tex
% ================================================================
\input{_tex/0_abstract}
\input{_tex/1_intro}

\input{_tex/2_background}
\input{_tex/3_analysis}
\input{_tex/4_method}
\input{_tex/5_experiment}
\input{_tex/6_conclusion}

% ================================================================
% reference
% ================================================================
\clearpage
\balance
\bibliographystyle{IEEEtran}
\bibliography{ref/reference}

\end{document}

%% file: _tex/0_abstract.tex
\begin{abstract}

VLA models have emerged as the mainstream for embodied intelligence, which decouples reasoning and execution via action chunks. 
To enable real-time deployment of VLA models, existing studies investigate edge-cloud collaborative inference, which pose two challenges to the action chunk mechanism.
(1) Edge-cloud collaboration complicates system latency components, hindering the determination of valid ranges for action chunks. 
(2) Latency fluctuations in cross-edge update links under edge-cloud scenarios make it difficult to schedule replanning requests.
To address these, we analyze the key factors affecting action chunk usable ranges and replanning request timing, revealing phase dependence and edge-cloud latency constraints.
Based on these insights, we propose \textit{ActECC}, which enables Action Chunk Temporal Harnessing through kinematics-guided usable range assessment and delay-aware replanning scheduling.
Experiments show that \textit{ActECC} reduces per-action latency by up to 27.4\% and cloud calls by up to 45.1\%, while lowering average execution error compared with existing edge-cloud collaborative systems.

\end{abstract}

%% file: _tex/1_intro.tex
\vspace{-2mm}
\section{\textbf{Introduction}}
\label{sec:intro}

% ===============================================================
% subsection 1: VLA背景 + Action Chunk背景
% ===============================================================
Vision-language-action (VLA) models represent a dominant paradigm for embodied intelligence due to their strong performance~\cite{roboflamingo,openvla,pi0,pi05,octo,rdt1b,tinyvla}.
As technology advances, VLA models employ \underline{action chunking} to lower inference frequency by generating a sequence of consecutive actions in one forward pass for sequential execution~\cite{act,diffusionpolicy,rtc,OpenVLA-OFT}.
Consequently, action chunking requires temporal coordination between the execution of the current action chunk and the generation of the next~\cite{SmolVLA,CheckVLA}.

% ===============================================================
% subsection 2: 端云
% ===============================================================
VLA incurs substantial computational overhead and hinders real-time deployment~\cite{spvla,SpecPrune,KERV,dyq}. Therefore, edge-cloud collaboration (ECC) serves as an effective deployment scheme for VLA, which offloads most parameters and computations to the cloud to accelerate action generation under constrained edge computing resources.
Recent studies investigate remote guidance with onboard adaptation~\cite{asyncvla}, intent- and network-aware partitioning~\cite{avery}, and fresh local visual observations~\cite{cloudedgevla}.

% ===============================================================
% subsection 3: challenge
% ===============================================================
However, edge-cloud collaboration introduces two coupled temporal challenges for action-chunk execution. 
\textbf{Challenge~\ding{172}:} The usable execution range of a chunk varies with robot motion, making fixed chunk lengths difficult to balance execution error and update overhead. 
\textbf{Challenge~\ding{173}:}
Variable computation and communication delays make it difficult to determine when to request the next chunk: early requests may waste resources by replacing still-usable actions, whereas late requests may leave execution gaps. 
Addressing these challenges requires coordinating action validity with the time needed to generate and deploy new actions.

\input{fig/challenge}

% ===============================================================
% subsection 4: analysis
% ===============================================================
% 为了解决上述两种挑战 xxx，我们进行了两项在端云下VLA执行系统的分析，这里也不提 Action Chink，是为了分析两个端云场景引发的VLA 执行系统引发问题。

% 需要凝练一下我们为什么分析，分析了什么，得到什么 insight，不需要太细节的说 xxx 指标啥的。
To address these challenges, we conduct two analyses of action-chunk execution in edge-cloud collaboration systems. 
\textbf{Analysis \ding{172}:} 
% First, we examine how motion and contact indicators reflect kinematics-guided execution tolerance.
% We then relate within-phase candidate disagreement to usable boundaries. Comparing phase-only, disagreement-only and combined assessments tests whether phase information helps interpret disagreement for low-cost usable range distributions.
First, we examine how variable edge-cloud delays challenge fixed action-chunk usable ranges.
We then investigate kinematic indicators, revealing their temporal-dependent relationship with usable ranges to guide adaptive determination.
\textbf{Analysis \ding{173}:} 
% Second, with boundary assessment fixed, we compare request strategies under similar mean execution and update times but different variability, examining execution constraints, waiting margins and trade-offs between lateness and cloud calls to find the latest feasible timing.
Second, we examine how variable edge-cloud update times complicate replanning request timing.
We then analyze pipeline parallelism, revealing its role in guiding request timing to reduce computation–execution gaps while preserving task accuracy.

% ===============================================================
% subsection 5: methods
% ===============================================================
% 这里提一下 Action Chunk Temporal Harnessing 并解释，我们的方法是xxx 的方法
% We propose \textit{ActECC}, an edge-cloud VLA inference framework implementing Action Chunk Temporal Harnessing through two edge-side mechanisms.

Based on the above analysis, we propose \textit{ActECC}, an edge-cloud collaborative VLA inference framework for action chunk harnessing.
The framework incorporates two core methods:
\textbf{Method \ding{172}:} action-aware usable range determination, which uses local kinematic indicators to dynamically estimate action-chunk usable ranges; 
\textbf{Method \ding{173}:} network-aware replanning scheduling, which uses the remaining usable execution window and estimated end-to-end update latency to select the latest feasible replanning request time.
Together, these mechanisms aim to reduce redundant cloud updates and pipeline bubbles while preserving task performance. Our contributions are below:

% ===============================================================
% subsection 6: contribution
% ===============================================================

\input{fig/analysis}

\begin{itemize}[leftmargin=*]
\item We explicitly analyze action-chunk execution in edge-cloud VLA systems, revealing the roles of kinematic indicators and pipeline parallelism.
\item We propose \textit{ActECC}, an action chunk harnessing framework for VLA's ECC deployment, through chunk usable range determination and chunk replanning scheduling.
\item We conduct detailed simulations and real-world evaluations to demonstrate the effectiveness and superiority of \textit{ActECC}.
\end{itemize}

%% file: fig/challenge.tex
\begin{figure}[t]
 \centering
 \includegraphics[width=3.3in,height=2in]{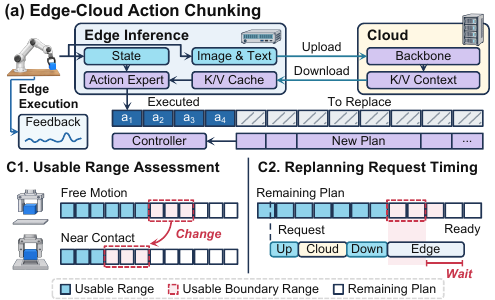}
 \vspace{-3mm}
 \caption{
 Edge-Cloud Deployments Posing Challenges on Action Chunking
 }
 \vspace{-4mm}
 \label{fig:timing}
\end{figure}

%% file: fig/analysis.tex
\begin{figure*}[!t]
 \centering
 \includegraphics[width=7in]{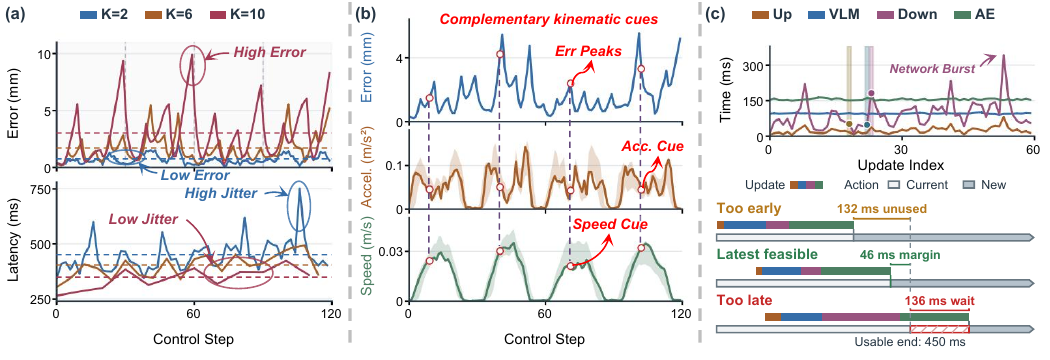}
 \vspace{-4mm}
 \caption{Action-Chunk Analysis: (a) Fixed-Range Trade-Offs; (b) Complementary Kinematic Cues; (c) Replanning Timing.}
 \vspace{-4mm}
 \label{fig:design}
\end{figure*}

%% file: _tex/2_background.tex
\section{\textbf{Background and Related Works}}
\label{sec:background}

%==============================================================
% subsection 1: VLA Inference System and Action Chunk Mechanism
% ==============================================================
% 其余的部分，应该尽可能对齐一下标题的关键词，特别是 System，Mechanism这种词。
\subsection{\textbf{VLA Inference System and Action Chunk Mechanism}}
\label{sec:background_1}

VLA inference systems generate discrete action tokens autoregressively~\cite{fast}, as in OpenVLA~\cite{openvla}, or continuous action chunks through flow-matching experts, as in $\pi_{0.5}$~\cite{pi05} and GR00T N1.5~\cite{grootn15}.
To reduce inference frequency, action chunking~\cite{act,diffusionpolicy,adachunk,vlacorrector} generates action sequences and executes a prefix before replanning with updated observations.
This amortizes inference costs across control steps; prediction and execution lengths can differ~\cite{moh,vlaknows,pace}.

%==============================================================
% subsection 2: Edge-Cloud Collaborative Inference System for VLA
% ==============================================================
\subsection{\textbf{Edge-Cloud Collaborative Inference System for VLA}}
\label{sec:background_2}

Edge-cloud VLA systems offload intensive computation to the cloud while retaining action execution at the edge~\cite{ecovla}.
For instance, RoboECC~\cite{roboecc} partitions VLA computation.
AsyncVLA~\cite{asyncvla} combines remote guidance with local adaptation, whereas CloudEdgeVLA~\cite{cloudedgevla} integrates fresh local observations.
% 凝练一点，强化 Challenge 即可
% During execution, cloud computation, communication, and edge processing prepare the next update.
% The current chunk must remain usable until new actions are ready.
% Its usable range varies with task phase and physical state, while update latency fluctuates.
However, variable edge-cloud delays complicate temporal coordination of action-chunk execution, particularly usable range determination and replanning request timing.

%% file: _tex/3_analysis.tex
\section{\textbf{Analysis}}
\label{sec:analysis}

%==============================================================
% subsection 1: Usable Range of Action Chunks under ECC Deployment
% ==============================================================
\subsection{\textbf{Usable Range of Action Chunks under ECC Deployment}}
\label{sec:analysis_1}

To investigate the limitations of fixed action-chunk execution ranges in edge-cloud collaborative VLA inference systems, we vary the number of actions executed between updates and measure execution errors and edge-cloud update latency (Fig.~\ref{fig:design}~(a)).
As the execution range increases, the average execution error rises, whereas shorter ranges require more frequent updates.
Meanwhile, execution errors vary considerably across control steps, and edge-cloud update latency exhibits noticeable fluctuations.
These observations indicate that fixed execution ranges cannot consistently balance action accuracy and update overhead, making it difficult to determine appropriate usable ranges for action chunks under ECC deployment.

To further investigate the factors affecting action-chunk usable ranges, we align execution errors with the robot's velocity and acceleration over the same control steps (Fig.~\ref{fig:design}~(b)).
We observe distinct motion patterns, with several error peaks occurring around periods of pronounced changes in velocity or acceleration.
Although neither indicator alone determines execution error, their temporal variations provide local cues for assessing the usable range of action chunks.
Therefore, we investigate cumulative local kinematic variations to dynamically estimate the remaining usable range under a calibrated task-specific deviation tolerance.

\textit{\textbf{Insight \ding{172}:}
Kinematic properties could help determine how long an action chunk remains usable.}

%==============================================================
% subsection 2: Chunks' Replanning Request under ECC Deployment
% ==============================================================
% \subsection{\textbf{Chunks' Replanning Request under ECC Deployment}}
% \label{sec:analysis_2}

% The usable range defines the execution window, while the update pipeline determines when new actions become available.
% However, their varying timing complicates periodic replanning.
% Early requests may replace still-usable actions and incur unnecessary updates, whereas late requests can create pipeline bubbles, leaving execution waiting for new actions (Fig.~\ref{fig:design}~(c)).

% We therefore quantify pipeline parallelism as the fraction of update time covered by usable action execution.
% We then compare request positions in terms of parallelism, pipeline bubbles, update overhead, and task performance.
% In particular, comparisons at similar parallelism examine whether later requests avoid premature updates while preserving timely handover (Fig.~\ref{fig:design}~(d)).

% \textit{\textbf{Insight \ding{173}:}
% Pipeline parallelism estimation guides replanning request timing to reduce computation–execution gaps while preserving task accuracy.}

\vspace{-1mm}
\subsection{\textbf{Chunk Replanning Request under ECC Deployment}}
\label{sec:analysis_2}

After determining the usable range of an action chunk, we further examine whether new actions can be generated before the current chunk expires.
To this end, we profile the latency of each edge-cloud update stage, including uplink transmission, cloud inference, downlink transmission, and edge action generation (Fig.~\ref{fig:design}~(c)).
We observe noticeable latency fluctuations across these stages, particularly occasional network delay spikes.
Consequently, periodic replanning requests may fail to ensure timely action updates.

To examine request timing, we compare early, timely, and late requests under the same usable execution window (Fig.~\ref{fig:design}~(c)).
With valid updates installed upon arrival, early requests may prematurely replace usable actions, whereas late requests can create execution bubbles.
In contrast, timely requests overlap updates with ongoing execution while preserving sufficient time for action handover.

Based on these observations, we quantify pipeline parallelism as the fraction of update time covered by usable action execution.
Insufficient parallelism may introduce execution gaps, whereas both early and timely requests can achieve full overlap.
Therefore, maximizing parallelism alone cannot determine the appropriate request timing.
Instead, replanning requests should be delayed as much as possible while ensuring timely action updates, thereby avoiding premature replacements and execution bubbles.

\textit{\textbf{Insight \ding{173}:}
The usable execution window and end-to-end update latency determine the latest feasible replanning request time with a calibrated safety margin.}

%% file: _tex/4_method.tex
\section{\textbf{\textit{ActECC} Framework}}
\label{sec:method}
\input{fig/method}

\textit{ActECC} framework implements action chunk temporal harnessing via two edge mechanisms: action-aware determination estimates the usable ranges of chunks, and network-aware scheduling selects the latest feasible replanning request time.

\subsection{\textbf{Action-Aware Chunk Usable Ranges Determination}}
\label{sec:method_1}
We define the usable range as the longest remaining prefix whose end-effector deviation from a timely replanned reference trajectory stays within a chosen threshold.
As illustrated in Fig.~\ref{fig:method}~(b), we estimate this range using cumulative motion derived from velocity and acceleration as a calibrated proxy for trajectory deviation.
At each control step, we estimate the remaining usable execution length as shown in Eq.~\eqref{eq:usable_range}.
\begin{equation}
\begin{aligned}
\widehat h_t
= \max_{h \in \{0,\ldots,H_t\}} \biggl\{ 
B_t + \sum_{j=0}^{h-1}
\left(
\|\widehat{\mathbf v}_{t,j}\|\Delta
+ \lambda_a\|\widehat{\mathbf a}_{t,j}\|\Delta^2
\right)
\leq \delta
\biggr\}.
\end{aligned}
\label{eq:usable_range}
\end{equation}
Here, $H_t$ denotes the number of remaining actions, $\Delta$ the control period, and $\widehat{\mathbf v}_{t,j}$, $\widehat{\mathbf a}_{t,j}$ the predicted velocity and acceleration at future control step $j$. 
The coefficient $\lambda_a$ weights the acceleration term. 
$B_t$ tracks the kinematic variation accumulated from the observation for generating the current chunk, with $\delta$ as its empirical budget. 
We calibrate $\lambda_a$ and $\delta$ on separate deferred-update records by matching estimated prefixes to the measured deviation boundaries.

During execution, local feedback updates the accumulated contribution of executed actions, while the remaining actions provide estimates for future steps.
The budget is therefore progressively consumed from a fixed reference observation and is not reset at each control step.
This also accounts for actions executed during chunk generation, maintaining a consistent reference as execution advances.
We retain the longest prefix satisfying the budget and obtain the remaining usable time as $\widehat T_t=\widehat h_t\Delta$.
If the budget is exhausted, the remaining usable range is set to zero.
Thus, action-aware determination translates local kinematic indicators into a dynamic usable range, providing the execution window for subsequent replanning.

\subsection{\textbf{Network-Aware Chunk Replanning Scheduling}}
\label{sec:method_2}
Based on the above insight, we adopt pipeline parallelism to schedule replanning requests for VLA inference in ECC deployment. 
The usable range specifies the remaining execution window, while the update pipeline governs the availability timing of new actions. 
We thereby estimate the total update duration $\widehat D_t$ along the dependency path, covering data transfer, cloud computation, and edge generation and deployment. 
The edge node quantifies this duration from request issuance to action readiness via a unified monotonic clock. 
Stage profiles are refreshed with recent measurements, where newer observations receive higher weights to capture runtime variations.

We quantify pipeline parallelism as the fraction of update time covered by usable action execution for each candidate request.
As illustrated in Fig.~\ref{fig:method}~(c), late requests may create execution bubbles, while timely requests enable full overlap between computation and execution.
Therefore, we select the latest feasible request to utilize the current chunk while retaining a time buffer, as shown in Eq.~\eqref{eq:replanning_request}.
\begin{equation}
w_t^\star
=
\max_{w\in\mathcal W_t}
\left\{
w:
w+\widehat D_t+m_t\leq\widehat T_t
\right\}.
\label{eq:replanning_request}
\end{equation}
Here, $w$ denotes the scheduled waiting time, and $\mathcal W_t$ contains admissible request positions.
The safety buffer $m_t \geq 0$ accounts for usable-time overestimation and update-duration underestimation.
It is calibrated using an upper empirical quantile of their combined estimation errors from independent execution records and completed updates.

At each control step, updated $\widehat T_t$ and $\widehat D_t$ adjust request timing, advancing replanning as the usable range shrinks or update latency grows.
A single in-flight update overlaps execution; Upon arrival, valid updates replace the uncommitted suffix after timestamp alignment, without waiting for the current chunk to expire; committed actions are preserved and stale actions discarded.
If no feasible request time exists, replanning starts immediately, followed by waiting or recovery upon usable range expiration.
Thus, network-aware scheduling maintains computation–execution overlap.

%% file: fig/method.tex
\begin{figure*}[!t]
 \centering
 \includegraphics[width=7in]{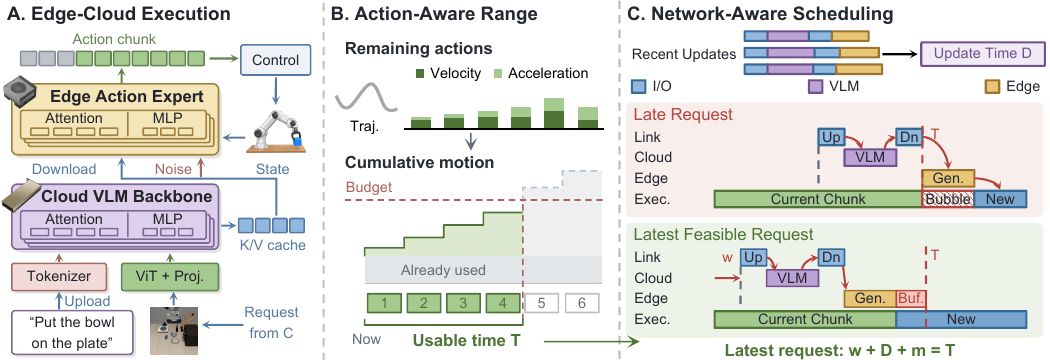}
 \vspace{-3mm}
 \caption{Overview of \textit{ActECC}: (a) Edge-Cloud Execution; (b) Action-Aware Range Estimation; (c) Network-Aware Request Scheduling.}
 \vspace{-4mm}
 \label{fig:method}
\end{figure*}

%% file: _tex/5_experiment.tex
\vspace{-1mm}
\section{\textbf{Experiments}}
\label{sec:experiments}

\input{table/experiment_style}

\input{table/main_pi05}

\input{fig/exp}

\vspace{-1.5mm}
\subsection{\textbf{Setup}}
We evaluate \textit{ActECC} on $\pi_{0.5}$~\cite{pi05} and GR00T N1.5~\cite{grootn15} across the LIBERO benchmark~\cite{libero} (Spatial, Object, Goal, and Long suites).
Baselines include edge-only, cloud-only, and fixed-length edge--cloud execution.
We use Jetson AGX Orin and Thor as edge devices, paired with a cloud NVIDIA A100 GPU.
We use execution error (Err., mm) against aligned human-expert demonstration trajectories rather than success rate (SR) to capture fine-grained execution deviations.
We also report amortized per-action latency (Lat., ms), the number of cloud calls per episode (Calls), and peak edge GPU memory usage (Load, GB).
The total runtime overhead includes both data preparation and runtime orchestration.

\vspace{-1.5mm}
\subsection{\textbf{Evaluation Results}}

\textbf{Simulation Results.}
Tabs.~\ref{tab:main} and~\ref{tab:main-groot} report the execution accuracy and system performance of \textit{ActECC} across VLA models and hardware platforms.
The results show that \textit{ActECC} achieves the lowest execution error across all four LIBERO suites.
Compared with Fixed ECC, \textit{ActECC} reduces average error by 7.9\% and 7.0\% on Orin and Thor with $\pi_{0.5}$, and by 7.7\% and 6.5\% with GR00T N1.5, respectively.
Moreover, \textit{ActECC} reduces edge GPU memory usage by 69.8--78.4\% compared with Edge-Only and per-action latency by 18.8--27.4\% compared with Fixed ECC.
These results show that \textit{ActECC} improves execution accuracy while reducing cloud calls by 31.3--45.1\% relative to Fixed ECC.

\textbf{Real-World Evaluations.}
We further evaluate \textit{ActECC} with $\pi_{0.5}$ on an Agilex Piper-X arm.
Fig.~\ref{fig:realworld} shows edge--cloud interactions between successive states of an orange-placement and lid-closing task.
Each update combines cloud backbone computation with edge-side action generation.
\textit{ActECC} estimates the longest usable range and schedules timely replanning, maintaining low error while reducing cloud computation.

\input{table/main_groot}
\input{table/ablation_results}
\input{table/overhead}

\vspace{-2mm}
\subsection{\textbf{Discussion}}
\textbf{Ablation Study.}
Tab. III evaluates the two mechanisms of ActECC using separate runs with $\pi_{0.5}$ on Jetson AGX Thor.
Range determination alone reduces execution error by 3.6\% and cloud calls by 12.6\%, while request scheduling alone reduces error by 1.8\%.
Combining both reduces execution error by 7.0\% to 61.4 mm and cloud calls by 45.1\% relative to the baseline.
These results highlight their complementary roles in execution accuracy and update coordination.

\textbf{Runtime Overhead.}
Tab.~\ref{tab:overhead} reports the runtime overhead across hardware platforms.
The total overhead is 310.19 $\mu$s on Orin, 106.82 $\mu$s on Thor, and 122.62 $\mu$s on the A100 Host.
Range determination dominates the computation, while scheduling incurs only 13.77--32.19 $\mu$s on edge devices.
Sub-millisecond overhead enables lightweight runtime deployment.

%% file: table/experiment_style.tex
% 仅定义实验表格配色。
\definecolor{ExpTitle}{HTML}{244DA5}
\definecolor{ExpEdge}{HTML}{DCDCF5}
\definecolor{ExpCloud}{HTML}{F7D5D8}
\definecolor{ExpFixed}{HTML}{FCE5CD}
\definecolor{ExpOurs}{HTML}{D8EED8}
\definecolor{ExpPartial}{HTML}{F0F0F0}

%% file: table/main_pi05.tex
\begin{table}[t]
\centering
\caption{Evaluation Results on LIBERO Benchmark with $\pi_{0.5}$}
\vspace{-1mm}
\label{tab:main}
\scriptsize
\setlength{\tabcolsep}{1.3mm}

\begin{tabularx}{\columnwidth}{c|c|*{4}{>{\centering\arraybackslash}X}|ccc}
\toprule\toprule
\multirow{2}{*}{\textbf{HW}} & \multirow{2}{*}{\textbf{Method}} & \multicolumn{4}{c|}{\textbf{Err. (mm)}} & \multicolumn{3}{c}{\textbf{System Metrics}} \\
\cmidrule{3-9}
 & & \textbf{Spatial} & \textbf{Object} & \textbf{Goal} & \textbf{Long} & \textbf{Lat.} & \textbf{Calls} & \textbf{Load} \\
\midrule
\multirow{4}{*}{\textbf{Orin}} & \cellcolor{ExpEdge}Edge-Only & \cellcolor{ExpEdge}73.6 & \cellcolor{ExpEdge}76.2 & \cellcolor{ExpEdge}68.9 & \cellcolor{ExpEdge}87.4 & \cellcolor{ExpEdge}612.40 & \cellcolor{ExpEdge}0.00 & \cellcolor{ExpEdge}7.645 \\
 & \cellcolor{ExpCloud}Cloud-Only & \cellcolor{ExpCloud}63.1 & \cellcolor{ExpCloud}65.7 & \cellcolor{ExpCloud}58.3 & \cellcolor{ExpCloud}75.1 & \cellcolor{ExpCloud}179.74 & \cellcolor{ExpCloud}22.01 & \cellcolor{ExpCloud}0.000 \\
 & \cellcolor{ExpFixed}Fixed ECC & \cellcolor{ExpFixed}66.4 & \cellcolor{ExpFixed}70.1 & \cellcolor{ExpFixed}62.2 & \cellcolor{ExpFixed}80.9 & \cellcolor{ExpFixed}473.60 & \cellcolor{ExpFixed}21.41 & \cellcolor{ExpFixed}1.654 \\
 & \cellcolor{ExpOurs}\textbf{\textit{ActECC}} & \cellcolor{ExpOurs}\textbf{61.7} & \cellcolor{ExpOurs}\textbf{64.8} & \cellcolor{ExpOurs}\textbf{56.9} & \cellcolor{ExpOurs}\textbf{74.2} & \cellcolor{ExpOurs}\textbf{384.80} & \cellcolor{ExpOurs}\textbf{12.32} & \cellcolor{ExpOurs}\textbf{1.654} \\
\midrule
\multirow{4}{*}{\textbf{Thor}} & \cellcolor{ExpEdge}Edge-Only & \cellcolor{ExpEdge}64.2 & \cellcolor{ExpEdge}66.5 & \cellcolor{ExpEdge}59.8 & \cellcolor{ExpEdge}76.4 & \cellcolor{ExpEdge}295.56 & \cellcolor{ExpEdge}0.00 & \cellcolor{ExpEdge}7.645 \\
 & \cellcolor{ExpCloud}Cloud-Only & \cellcolor{ExpCloud}63.1 & \cellcolor{ExpCloud}65.7 & \cellcolor{ExpCloud}58.3 & \cellcolor{ExpCloud}75.1 & \cellcolor{ExpCloud}179.74 & \cellcolor{ExpCloud}22.01 & \cellcolor{ExpCloud}0.000 \\
 & \cellcolor{ExpFixed}Fixed ECC & \cellcolor{ExpFixed}63.8 & \cellcolor{ExpFixed}65.9 & \cellcolor{ExpFixed}58.7 & \cellcolor{ExpFixed}75.6 & \cellcolor{ExpFixed}240.02 & \cellcolor{ExpFixed}21.41 & \cellcolor{ExpFixed}1.654 \\
 & \cellcolor{ExpOurs}\textbf{\textit{ActECC}} & \cellcolor{ExpOurs}\textbf{59.4} & \cellcolor{ExpOurs}\textbf{61.9} & \cellcolor{ExpOurs}\textbf{54.4} & \cellcolor{ExpOurs}\textbf{69.8} & \cellcolor{ExpOurs}\textbf{186.28} & \cellcolor{ExpOurs}\textbf{11.76} & \cellcolor{ExpOurs}\textbf{1.654} \\
\bottomrule\bottomrule
\end{tabularx}
\vspace{-4mm}
\end{table}

%% file: fig/exp.tex
\begin{figure*}[!b]
 \centering
 \vspace{-5mm}
 \includegraphics[width=7in]{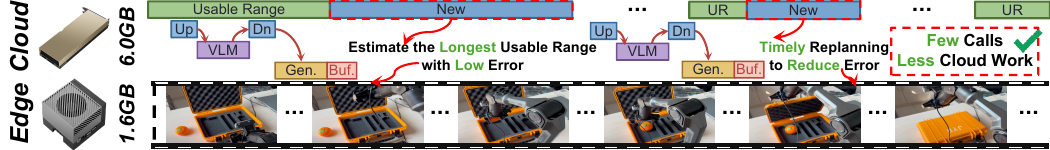}
 \vspace{-3mm}
 \caption{A Case of Real-World Edge-Cloud VLA Execution with \textit{ActECC} (Task: Place an Orange in a Box and Close the Box).}
 \label{fig:realworld}
\end{figure*}

%% file: table/main_groot.tex
\begin{table}[t]
\centering
\caption{Evaluation Results on LIBERO Benchmark with GR00T N1.5}
\label{tab:main-groot}
\vspace{-1mm}
\scriptsize
\setlength{\tabcolsep}{1.5mm}
\begin{tabularx}{\columnwidth}{c|c|*{4}{>{\centering\arraybackslash}X}|ccc}
\toprule
\toprule
\multirow{2}{*}{\textbf{HW}} & \multirow{2}{*}{\textbf{Method}} & \multicolumn{4}{c|}{\textbf{Err. (mm)}} & \multicolumn{3}{c}{\textbf{System Metrics}} \\
\cmidrule{3-9}
 & & \textbf{Spatial} & \textbf{Object} & \textbf{Goal} & \textbf{Long} & \textbf{Lat.} & \textbf{Calls} & \textbf{Load} \\
\midrule
\multirow{4}{*}{\textbf{Orin}} & \cellcolor{ExpEdge}Edge-Only & \cellcolor{ExpEdge}103.7 & \cellcolor{ExpEdge}108.2 & \cellcolor{ExpEdge}96.8 & \cellcolor{ExpEdge}122.3 & \cellcolor{ExpEdge}307.64 & \cellcolor{ExpEdge}0.00 & \cellcolor{ExpEdge}6.945 \\
 & \cellcolor{ExpCloud}Cloud-Only & \cellcolor{ExpCloud}92.4 & \cellcolor{ExpCloud}96.1 & \cellcolor{ExpCloud}84.7 & \cellcolor{ExpCloud}106.8 & \cellcolor{ExpCloud}137.11 & \cellcolor{ExpCloud}32.69 & \cellcolor{ExpCloud}0.000 \\
 & \cellcolor{ExpFixed}Fixed ECC & \cellcolor{ExpFixed}93.8 & \cellcolor{ExpFixed}99.1 & \cellcolor{ExpFixed}87.6 & \cellcolor{ExpFixed}113.5 & \cellcolor{ExpFixed}235.44 & \cellcolor{ExpFixed}31.80 & \cellcolor{ExpFixed}2.100 \\
 & \cellcolor{ExpOurs}\textbf{\textit{ActECC}} & \cellcolor{ExpOurs}\textbf{87.4} & \cellcolor{ExpOurs}\textbf{91.3} & \cellcolor{ExpOurs}\textbf{80.2} & \cellcolor{ExpOurs}\textbf{104.8} & \cellcolor{ExpOurs}\textbf{170.97} & \cellcolor{ExpOurs}\textbf{21.21} & \cellcolor{ExpOurs}\textbf{2.100} \\
\midrule
\multirow{4}{*}{\textbf{Thor}} & \cellcolor{ExpEdge}Edge-Only & \cellcolor{ExpEdge}90.5 & \cellcolor{ExpEdge}94.3 & \cellcolor{ExpEdge}84.2 & \cellcolor{ExpEdge}107.1 & \cellcolor{ExpEdge}196.50 & \cellcolor{ExpEdge}0.00 & \cellcolor{ExpEdge}6.945 \\
 & \cellcolor{ExpCloud}Cloud-Only & \cellcolor{ExpCloud}92.4 & \cellcolor{ExpCloud}96.1 & \cellcolor{ExpCloud}84.7 & \cellcolor{ExpCloud}106.8 & \cellcolor{ExpCloud}137.11 & \cellcolor{ExpCloud}32.69 & \cellcolor{ExpCloud}0.000 \\
 & \cellcolor{ExpFixed}Fixed ECC & \cellcolor{ExpFixed}89.7 & \cellcolor{ExpFixed}93.2 & \cellcolor{ExpFixed}82.9 & \cellcolor{ExpFixed}105.8 & \cellcolor{ExpFixed}190.03 & \cellcolor{ExpFixed}31.80 & \cellcolor{ExpFixed}2.100 \\
 & \cellcolor{ExpOurs}\textbf{\textit{ActECC}} & \cellcolor{ExpOurs}\textbf{84.1} & \cellcolor{ExpOurs}\textbf{87.7} & \cellcolor{ExpOurs}\textbf{77.0} & \cellcolor{ExpOurs}\textbf{98.5} & \cellcolor{ExpOurs}\textbf{142.27} & \cellcolor{ExpOurs}\textbf{21.86} & \cellcolor{ExpOurs}\textbf{2.100} \\
\bottomrule\bottomrule
\end{tabularx}
\vspace{-4mm}
\end{table}

%% file: table/ablation_results.tex
\begin{table}[!tb]
\centering
\caption{Ablation Study about the \textit{ActECC} Framework}
\vspace{-1mm}
\label{tab:ablation}
\scriptsize
\setlength{\tabcolsep}{1.5mm}
\begin{tabularx}{\columnwidth}{>{\centering\arraybackslash}X>{\centering\arraybackslash}X|*{4}{>{\centering\arraybackslash}X}}
\toprule\toprule
\multicolumn{2}{c|}{\textbf{Mechanisms}} & \multicolumn{4}{c}{\textbf{Task and System Performance}} \\
\cmidrule{1-2}\cmidrule{3-6}
\textbf{Range} & \textbf{Request} & \textbf{Err. (mm)} & \textbf{Lat.} & \textbf{Calls} & \textbf{Load} \\
\midrule
\rowcolor{ExpFixed}\ding{55} & \ding{55} & 66.0 & 280.02 & 21.41 & 1.654 \\
\rowcolor{ExpPartial}\ding{51} & \ding{55} & 63.6 & 210.87 & 18.72 & 1.654 \\
\rowcolor{ExpPartial}\ding{55} & \ding{51} & 64.8 & 225.28 & 17.54 & 1.654 \\
\rowcolor{ExpOurs}\ding{51} & \ding{51} & 61.4 & 184.47 & 11.76 & 1.654 \\
\bottomrule\bottomrule
\end{tabularx}
\vspace{-4mm}
\end{table}

%% file: table/overhead.tex
\begin{table}[!t]
\centering
\caption{Runtime Overhead of \textit{ActECC}}
\vspace{-1mm}
\label{tab:overhead}
\scriptsize
\setlength{\tabcolsep}{1mm}
\begin{tabularx}{\columnwidth}{@{}c|c|*{3}{>{\centering\arraybackslash}X}@{}}
\toprule
\toprule
 & \textbf{Hardware}
 & \textbf{Range}
 & \textbf{Schedule}
 & \textbf{Total} \\
\midrule
\multirow{2}{*}{Edge-Side}
 & Orin
 & \cellcolor{ExpOurs}242.72\,(0.49\%)
 & \cellcolor{ExpOurs}32.19\,(0.06\%)
 & \cellcolor{ExpOurs}310.19\,(0.62\%) \\
 & Thor
 & \cellcolor{ExpOurs}90.05\,(0.18\%)
 & \cellcolor{ExpOurs}13.77\,(0.03\%)
 & \cellcolor{ExpOurs}106.82\,(0.21\%) \\
\cmidrule{2-5}
Cloud-Side
 & A100 Host
 & \cellcolor{ExpOurs}98.27\,(0.20\%)
 & \cellcolor{ExpOurs}12.41\,(0.02\%)
 & \cellcolor{ExpOurs}122.62\,(0.25\%) \\
\bottomrule
\bottomrule
\end{tabularx}
\vspace{-4mm}
\end{table}

%% file: _tex/6_conclusion.tex
\section{\textbf{Conclusion}}
We propose \textit{ActECC} to coordinate action-chunk execution and replanning in edge-cloud VLA systems.
By adapting usable ranges and request timing, \textit{ActECC} reduces execution error with low runtime overhead.